\documentclass[aps,showpacs,10pt,twocolumn]{revtex4}
\usepackage{epsf}

\begin{document}

\title{Faster quantum walk algorithm for the two dimensional spatial search}         
\author{Avatar Tulsi\\
        {\small Department of Physics, Indian Institute of Science, Bangalore-560012, India}}
\email{tulsi9@gmail.com}

\begin{abstract}
We consider the problem of finding a desired item out of $N$ items arranged on the sites of a two-dimensional lattice of size $\sqrt{N} \times \sqrt{N}$. The previous quantum walk based algorithms take $O(\sqrt{N}\log N)$ steps to solve this problem, and it is an open question whether the performance can be improved. We present a new algorithm which solves the problem in $O(\sqrt{N\log N})$ steps, thus giving an $O(\sqrt{\log N})$ improvement over the known algorithms. The improvement is achieved by controlling the quantum walk on the lattice using an ancilla qubit.   
\end{abstract} 


\maketitle

\section{Introduction}

Suppose we have $N$ items arranged on a two-dimensional lattice of size $\sqrt{N}\times \sqrt{N}$. Let the sites be labeled by their $x$ and $y$ coordinates as $|x,y\rangle$ for $x,y \in \{0,\ldots,\sqrt{N}-1\}$. In the quantum scenario, the coordinates label the basis states of an $N$ dimensional Hilbert space. Let $f(x,y)$ be a binary function which is $1$ if the item placed on $|x,y\rangle$ site satisfies certain properties (i.e. is a marked item $m$), else it is $0$. We assume that there is a unique marked item and let $|m\rangle = |x,y\rangle_{f(x,y)=1}$ denote the corresponding site or basis state. The two dimensional spatial search problem is to find $|m\rangle$ using minimum time steps, with the constraints that in one time step we can either examine the current site (i.e. compute $f(x,y)$ for the current site using an oracle query) or move to a neighboring site.

The straightforward application of Grover's search algorithm~\cite{grover} cannot be used to solve the problem faster than classical search as pointed out by Benioff~\cite{benioff}. Although it can find $|m\rangle$ using $O(\sqrt{N})$ oracle queries, between successive queries, it needs to perform a reflection about a superposition of all sites. This reflection takes $O(\sqrt{N})$ time steps, as in one time step we can only move to a neighboring site and we have to move across $\sqrt{N}$ sites in each direction of the lattice to perform a reflection.  Note that in the standard search problem, there is no restriction on the movement on lattice, and hence this reflection is not a hurdle. But for $2d$ spatial search, the total complexity becomes $O(\sqrt{N}\times \sqrt{N})=O(N)$ time steps, no better than brute-force searching.

Aaronson and Ambainis have shown that using a cleverly designed recursion of the quantum search algorithm, the $2d$ spatial search problem can be solved in $O(\sqrt{N}\log^{2}N)$ time steps~\cite{aaronson}. A better alternative is provided by the quantum walk search algorithms. They have been constructed for spatial search in any number of dimensions (see, for example, Refs.~\cite{randomwalk,CGnondirac,AKR,CG}). For the $2d$ spatial search problem, the discrete time algorithm by Ambainis, Kempe, Rivosh (AKR)~\cite{AKR} and the continuous time algorithm by Childs and Goldstone (CG)~\cite{CG} can do the job in $O(\sqrt{N}\log N)$ time steps. It is an open question whether the algorithms can be further improved, particularly whether the lower bound of $\Omega(\sqrt{N})$~\cite{optimal} can be achieved. Here, we give a positive answer to this question by presenting an improved algorithm that can solve the two-dimensional spatial search problem in $O(\sqrt{N\log N})$ time steps, thus giving an $O(\sqrt{\log N})$ improvement over the best known algorithms.

We present our results in the context of AKR's discrete time quantum walk algorithm, but the same can be applied to the continuous time quantum walk algorithm of CG. These quantum walk algorithms start with a uniform superposition of all sites and achieve a particular state, denoted by $|\alpha^{+}\rangle$ here, in $O(\sqrt{N\log N})$ time steps. The overlap of $|\alpha^{+}\rangle$ with $|m\rangle$ state is $\Theta(1/\sqrt{\log N})$, so that $O(\sqrt{\log N})$ rounds of quantum amplitude amplification~\cite{qaa} can be used to get the $|m\rangle$ state with constant probability. Hence, the total complexity of the algorithms is $O(\sqrt{N\log N}\times \sqrt{\log N}) = O(\sqrt{N} \log N)$. In the case of AKR's algorithm, the quantum walk search is analysed by reducing it to an instance of the abstract search algorithm, which is a generalization of Grover's search algorithm. 

We modify the quantum walk algorithms in a particular way so that the $|\alpha^{+}\rangle$ state, obtained after $O(\sqrt{N\log N})$ walk steps on the uniform superposition, has a significant overlap with the $|m\rangle$ state. Hence no rounds of quantum amplitude amplification are required by the new algorithm, and the time complexity remains $O(\sqrt{N\log N})$. As we show, this improvement is possible by controlling the quantum walk on the lattice in a clever way using an ancilla qubit. Our algorithm applies to any instance of abstract search algorithm, so it can also be used for improving the spatial search in higher dimensions. But there the improvement is only by a constant factor. 

The paper is organized as follows: In section II, we review the abstract search algorithm, presented by AKR, with the example of two-dimensional spatial search. Although our analysis follows AKR's paper, we use different notation for convenience. In section III, we present the \emph{controlled} quantum walk algorithm. We conclude the paper with some discussions in section IV. In the Appendix, we present analysis of the abstract search algorithm, which closely follows AKR's analysis (see section $7$ of ~\cite{AKR}) and uses the results presented there. The difference is a minor modification which is required for the new algorithm.


\section{Background}

Grover's search algorithm starts with an initial state $|s\rangle$, normally chosen to be uniform superposition of all the basis states. The algorithm drives it to the target state $|t\rangle$ by successively applying the reflection operators, $R_{t}=2|t\rangle \langle t|-I_{N}$ and $R_{s}=2|s\rangle \langle s|-I_{N}$, where $I_{N}$ is the $N$-dimensional identity operator. The $|t\rangle$ ($|s\rangle$) state is an eigenstate of reflection operator $R_{t}$ ($R_{s}$) with eigenvalue $1$, and all the states orthogonal to $|t\rangle$ ($|s\rangle$) have eigenvalue $-1$. It has been shown that applying the operator $U_{G}=R_{s}\cdot R_{t}$ on $|s\rangle$ rotates it in the two-dimensional subspace spanned by $|s\rangle$ and $|t\rangle$, and after $O(1/|\langle t|s\rangle|)$ iterations of $U_{G}$ we come very close to the $|t\rangle$ state. 

The abstract quantum search algorithm is a generalization of Grover's search algorithm, where the operator $R_{t}$ remains the same but $R_{s}$ gets replaced by a more general operator $U$. The $|s\rangle$ state is still required to be an eigenstate of $U$ with eigenvalue $1$, but the states orthogonal to $|s\rangle$ need not be its eigenstates with eigenvalue $-1$ as in the case of $R_{s}$. Also, $U$ is required to be a real operator (not necessarily a reflection operator) and not to have any other eigenstate with eigenvalue $1$ apart from $|s\rangle$. The abstract search algorithm iterates the operator $U_{A}=U\cdot R_{t}$ to get to the target state $|t\rangle$. 

To quantify the number of iterations of $U_{A}$ needed to get to the $|t\rangle$ state, we note that since $U$ is a real unitary matrix, its non-$\pm 1$ eigenvalues come in pairs of complex conjugate numbers $e^{\pm i\theta}$. The eigenstate corresponding to eigenvalue $1$ ($\theta = 0$) is $|s\rangle$, also denoted as $|\Phi_{0}\rangle$ here. Let the eigenstates corresponding to eigenvalue $-1$ ($\theta = \pi$) be denoted by $|\Phi_{k}\rangle$, $k=1,\ldots M$. Let $|\Phi_{j}^{\pm}\rangle$ denote all other eigenvectors with non-$\pm 1$ eigenvalues $e^{\pm i\theta_{j}}$. Then $|\Phi_{j}^{+}\rangle = |\Phi_{j}^{-}\rangle^{*}$ as $U$ is real. Let $a_{j}^{\pm}=\langle \Phi_{j}^{\pm}|t\rangle$, $a_{0}=\langle \Phi_{0}|t\rangle$ and $a_{k}=\langle \Phi_{k}|t\rangle$ be the expansion coefficients of $|t\rangle$ in the eigenbasis of $U$. Since $|t\rangle$ is a real vector, $a_{j}^{+}=(a_{j}^{-})^{*}$, and upto a global phase, $|\Phi_{j}^{\pm}\rangle$ can be chosen such that $a_{j}^{+}=a_{j}^{-}=a_{j}$. Similarly upto a global phase $|\Phi_{0}\rangle$ and $|\Phi_{k}\rangle$ can be chosen such that $a_{0},a_{k}$ are real. Thus
\begin{equation}
|t\rangle = a_{0}|\Phi_{0}\rangle +\sum_{j}a_{j}(|\Phi_{j}^{+}\rangle +|\Phi_{j}^{-}\rangle) +\sum_{k}a_{k}|\Phi_{k}\rangle. \label{abstractexpansion}
\end{equation}

To analyse the iteration of operator $U_{A}=U\cdot R_{t}$ on $|\Phi_{0}\rangle$, its eigenspectrum was determined by AKR. Though they have not explicitly considered the possibility when $U$ has an eigenspace with eigenvalue $-1$, their analysis can be easily generalized to that case. As it is crucial for the new algorithm, we have done this analysis in Appendix for completeness. For particular cases (spatial search is one of them), only two eigenvectors $|\pm \alpha\rangle$ of $U_{A}$ with the eigenvalues $e^{\pm i\alpha}$ are important, with the starting state $|\Phi_{0}\rangle$ almost completely spanned by them. Here $\alpha$ depends upon the eigenspectrum of $U$ as
\begin{equation}
\alpha = \Theta \left(\frac{a_{0}}{\sqrt{\sum_{j}\frac{a_{j}^{2}}{1-cos\theta_{j}}+\frac{A_{k}^{2}}{4}}}\right), \label{abstracteigenvalues}
\end{equation}
where $A_{k}=\sqrt{\sum_{k=1}^{M}a_{k}^{2}}$ is the projection of $|t\rangle$ on the $-1$-eigenspace. As shown by AKR, $|\Phi_{0}\rangle$ is close to $|\alpha^{-}\rangle = \frac{-i}{\sqrt{2}}(|\alpha\rangle-|-\alpha\rangle)$. Quantitatively
\begin{equation}
|\langle \Phi_{0}|\alpha^{-}\rangle| \geq 1-\Theta \left(\alpha^{4}\sum_{j}\frac{a_{j}^{2}}{a_{0}^{2}}\frac{1}{(1-cos\theta_{j})^{2}}\right) - \Theta\left(\frac{A_{k}^{2}\alpha^{4}}{a_{0}^{2}}\right). \label{abstractinitialoverlap}
\end{equation}
After iterating the operator $U_{A}$ for $T=\lceil \pi/2\alpha \rceil$ times on $|\alpha^{-}\rangle$, we come very close to the state $|\alpha^{+}\rangle = -i(e^{i\pi/2}|\alpha\rangle-e^{-i\pi/2}|-\alpha\rangle)/\sqrt{2} = (|\alpha\rangle+|-\alpha\rangle)/\sqrt{2}$. As shown by AKR, the quantity $|\langle t|\alpha^{+}\rangle|$ depends upon the eigenspectrum of $U$ as
\begin{equation}
|\langle t|\alpha^{+}\rangle|= \Theta \left(min\left(\frac{1}{\sqrt{\sum_{j}a_{j}^{2}cot^{2}\frac{\theta_{j}}{4}}},1\right)\right). \label{abstracttargetoverlap} 
\end{equation}

Consequently, any operator $U$ can be used in place of $R_{s}$ for quantum search if it satisfies the conditions for abstract search algorithm, i.e. it is a real operator with the initial state $|s\rangle$ as its unique eigenstate of eigenvalue $1$. Sometimes this flexibility is very useful. In the case of spatial search, there is a restriction that in one time step, we can move only to neighboring lattice sites. In this case, $U$ can be chosen such that it can be implemented in only one time step, whereas $R_{s}$ takes $\Theta(\sqrt{N})$ steps. For any $U$, we need to find its eigenspectrum, and the expansion coefficients of the target state in its eigenbasis, in order to analyse the algorithm using Eqns. (\ref{abstractexpansion}-\ref{abstracttargetoverlap}).


\subsection{Two-dimensional spatial search} 

We illustrate the abstract search algorithm with the specific example of two dimensional spatial search. AKR's algorithm attaches a $4$-dimensional coin space $\mathcal{H}_{c}$ to the Hilbert space $\mathcal{H}_{N}$ associated with $N$ lattice sites, and works in the joint Hilbert space $\mathcal{H}_{J}=\mathcal{H}_{c}\otimes \mathcal{H}_{N}$. The four basis states $d=0,1,2,3$ of $\mathcal{H}_{c}$ represent the four possible directions of movements on a two-dimensional lattice, i.e. $|\rightarrow\rangle, |\leftarrow\rangle, |\uparrow\rangle, |\downarrow\rangle$. Let $|u_{c}\rangle = \frac{1}{2}\sum_{d}|d\rangle$ be their uniform superposition and let $|u_{N}\rangle = \sum_{x,y}|x,y\rangle)/\sqrt{N}$ be the uniform superposition of all lattice sites. The initial state $|\Phi_{0}\rangle$ of AKR's algorithm is $|\Phi_{0}\rangle_{AKR} = |u_{c}\rangle |u_{N}\rangle$ which can be prepared in $2\sqrt{N}$ time steps. (For preparing $|u_{N}\rangle$, the idea is to start with a site $|0,0\rangle$, first spread the amplitude along $x$ axis in $\sqrt{N}$ steps, and then repeat the process for $y$-axis in another $\sqrt{N}$ steps~\cite{AKR}.)

The algorithm then iteratively applies the operator $U_{W} = W \cdot \bar{R}_{uc,m}$ on $|\Phi_{0}\rangle$. The operator $\bar{R}_{uc,m}=-R_{uc,m} = I_{4N}-2|u_{c},m\rangle\langle u_{c},m|$ is the negative of the reflection about the $|u_{c}\rangle|m\rangle$ state. It can be implemented in one time step by examining the lattice sites (in quantum superposition) using an oracle, and then applying $\bar{R}_{uc} =-R_{uc} = I_{4}-2|u_{c}\rangle\langle u_{c}|$ iff the site is the marked site $|m\rangle$. The walk operator $W$ is a product of two operators, coin flip $R_{uc}\otimes I_{N}$ and the moving step $S$. The coin flip acts only on the coin space but the moving step $S$ acts jointly on coin- and lattice-space as
\begin{eqnarray}
S:&|\rightarrow\rangle \otimes |x,y\rangle  \longrightarrow  |\leftarrow\rangle \otimes |x+1,y\rangle, \nonumber  \\
  &|\leftarrow\rangle \otimes |x,y\rangle  \longrightarrow  |\rightarrow\rangle \otimes |x-1,y\rangle, \nonumber\\
  &|\uparrow\rangle \otimes |x,y\rangle  \longrightarrow  |\downarrow\rangle \otimes |x,y+1\rangle, \nonumber\\
  &|\downarrow\rangle \otimes |x,y\rangle  \longrightarrow |\uparrow\rangle \otimes |x,y-1\rangle. \label{movingstep}
\end{eqnarray}
As $S$ involves movement only between neighboring sites, $|x\rangle \rightarrow |x\pm 1\rangle$ and $|y\rangle \rightarrow |y\pm 1\rangle$, it can be implemented in one time step. Hence $U_{W}$ can be implemented in $2$ time steps, one for $W = S\cdot (R_{uc}\otimes I_{N})$ and another for $\bar{R}_{uc,m}$.

AKR have shown that their algorithm is an instance of the abstract search algorithm. The operator $U_{W}$ is equivalent to $W\cdot R_{uc,m}$ upto a sign, making $|u_{c}\rangle|m\rangle$ the effective target state $|t\rangle$. The walk operator $W$ satisfies the required properties for the abstract search algorithm, within a particular subspace that is preserved by $U_{W}$. It is easy to check that $|\Phi_{0}\rangle$ is an eigenvector of $W$ with eigenvalue $1$. The other eigenvectors of $W$ are 
\begin{equation} 
|\Phi_{pq}\rangle = |v_{pq}\rangle |\chi_{p}\rangle |\chi_{q}\rangle\ ,\ p,q\in \{0,\ldots,\sqrt{N}-1\}, \label{AKReigenvectors}
\end{equation}
where $|\chi_{p}\rangle = \frac{1}{\sqrt[4]{N}}\sum_{x=0}^{\sqrt{N}-1}e^{i2\pi p\cdot x/\sqrt{N}}|x\rangle$ and $|\chi_{q}\rangle = \frac{1}{\sqrt[4]{N}}\sum_{y=0}^{\sqrt{N}-1}e^{i2\pi q\cdot y}|y\rangle$ form the Fourier basis. For each $p$ and $q$, there are four eigenvalues $1,-1$ and $e^{\pm i\theta_{pq}}$ with
\begin{equation}
cos\theta_{pq}=\frac{1}{2}\left(cos\frac{2\pi p}{\sqrt{N}}+cos\frac{2\pi q}{\sqrt{N}}\right), \label{AKReigenvalues}
\end{equation}
corresponding to four different vectors $|v_{pq}^{1}\rangle$, $|v_{pq}^{-1}\rangle$ and $|v_{pq}^{\pm}\rangle$ of the coin space. $W$ satisfies the conditions of the abstract search algorithm within the subspace $\mathcal{H}_{0}$ spanned by the eigenstates $|\Phi_{pq}^{\pm}\rangle =|v_{pq}^{\pm}\rangle|\chi_{p}\rangle|\chi_{q}\rangle$, $(p,q)\neq (0,0)$ and $|\Phi_{0}\rangle$, and $|\Phi_{0}\rangle$ is a unique eigenstate with eigenvalue $1$ within this subspace. AKR have shown that the operator $U_{W}$ preserves this subspace. 

As shown by AKR, the vectors $|v_{pq}^{\pm}\rangle$ are such that $a_{pq}=\langle \Phi_{pq}^{\pm}|u_{c},m\rangle=1/\sqrt{2N}$. We also have $a_{0}=\langle \Phi_{0}|u_{c},m\rangle = \langle u_{c}|u_{c}\rangle\langle u_{N}|m\rangle = 1/\sqrt{N}$. Using these values and Eq. (\ref{AKReigenvalues}) for $\theta_{pq}$, AKR have shown that the sums in Eqns. (\ref{abstracteigenvalues}-\ref{abstracttargetoverlap}) are 
\begin{eqnarray}
&\sum_{p,q}\frac{a_{pq}^{2}}{1-cos\theta_{pq}} = \Theta(\log N),& \label{AKRsum1}\\
&\sum_{p,q}\frac{\alpha^{4}}{a_{0}^{2}}\frac{a_{pq}^{2}}{(1-cos\theta_{pq})^{2}}= \Theta (\frac{1}{\log^{2}N}),& \label{AKRsum2}\\
&\sum_{p,q}a_{pq}^{2}cot^{2}\frac{\theta_{pq}}{4} = \Theta(\log N).& \label{AKRsum3}
\end{eqnarray}
Since the eigenstates $|v_{pq}^{-1}\rangle|\chi_{p}\rangle|\chi_{q}\rangle$ are orthogonal to $\mathcal{H}_{0}$, they do not matter for the algorithm and do not contribute to $A_{k}$. For even $\sqrt{N}$, there are two eigenstates $|\Phi_{\sqrt{N}/2,\sqrt{N}/2}^{\pm}\rangle$ of $W$ having eigenvalue $-1$ within $\mathcal{H}_{0}$. Since the projection of the target state on these eigenstates is $O(1/\sqrt{N})$, their contribution to $A_{k}$ is negligible. 

Putting above values in Eqns. (\ref{abstracteigenvalues}-\ref{abstracttargetoverlap}), we get
\begin{eqnarray}
&\alpha = \Theta(1/\sqrt{N\log N}),& \label{AKReigenvalue}\\
&|\langle \Phi_{0}|\alpha^{-}\rangle| \geq 1-\Theta (\frac{1}{\log^{2}N}),& \label{AKRinitial}\\
&|\langle u_{c},m|\alpha^{+}\rangle| = \Theta\left(\frac{1}{\sqrt{\log N}}\right).& \label{AKRtarget}
\end{eqnarray}
Hence, we have $|\Phi_{0}\rangle = |\alpha^{-}\rangle +|\epsilon\rangle$, with $\|\epsilon\| = \Theta(1/\log N)$. After $\lceil \pi/2\alpha\rceil = O(\sqrt{N\log N})$ quantum walk steps, the state becomes $|\alpha^{+}\rangle + |\epsilon'\rangle$, with $\|\epsilon'\|= \Theta(1/\log N)$. Since $|\langle \alpha^{+}|u_{c},m\rangle|$ is $\Theta(1/\sqrt{\log N})$ and $|\langle \epsilon'|u_{c},m\rangle|$ is of lower order $O(1/\log N)$, the overlap of the final state with $|u_{c},m\rangle$ is $\Theta(1/\sqrt{\log N})$. Thus we can get the $|u_{c},m\rangle$ state, or the $|m\rangle$ state, using $\sqrt{\log N}$ rounds of quantum amplitude amplification. The total number of time steps becomes $O(\sqrt{N\log N} \times \sqrt{\log N}) =O(\sqrt{N}\log N)$. 

In the next section, we show that by controlling the quantum step using an ancilla qubit, the coefficients $a_{pq}$, $a_{0}$ and $A_{k}$ can be manipulated in such a way that no rounds of quantum amplitude amplification are required and the $|m\rangle$ state can be obtained in $O(\sqrt{N\log N})$ time steps.


\section{Controlled quantum walk algorithm}
The new algorithm attaches an ancilla qubit $|b\rangle$ to the system, which controls the operations in the joint Hilbert space. The algorithm works in the $8N$-dimensional Hilbert space $\mathcal{H} = \mathcal{H}_{b}\otimes \mathcal{H}_{c}\otimes \mathcal{H}_{N}$, where $\mathcal{H}_{b}$ is the two-dimensional Hilbert space of the ancilla qubit. We use the subscripts $b$ and $J$ respectively for denoting the states or operations within the ancilla qubit space $\mathcal{H}_{b}$ and the joint Hilbert space $\mathcal{H}_{J} = \mathcal{H}_{c}\otimes \mathcal{H}_{N}$. (Note that $\mathcal{H}_{J}$ is the working space of AKR's algorithm.) 

The circuit diagram of the algorithm is shown in Fig. 1. The initial state is $|\Phi_{0}\rangle_{cqw} = |1\rangle|u_{c}\rangle|u_{N}\rangle = |1\rangle \otimes |\Phi_{0}\rangle_{AKR}$, and it can be prepared in $O(\sqrt{N})$ time steps. The controlled quantum walk algorithm then iteratively applies the operator 
\begin{equation}
U_{C} = (\bar{Z})_{b}\cdot c_{1}W \cdot (X_{\delta}^{\dagger})_{b} \cdot c_{1}\bar{R}_{uc,m} \cdot (X_{\delta})_{b} \label{definitionUC}
\end{equation}
to $|\Phi_{0}\rangle_{cqw}$. Note that in the figure, the operations are performed sequentially from left to right, while in equations they are performed from right to left. $X_{\delta}$ and $\bar{Z}$ are the single qubit gates given by
\begin{equation}
X_{\delta}= \left( \begin{array}{cc} cos\delta & sin\delta \\ -sin\delta & cos\delta\end{array} \right)\ ,\ \bar{Z}= \left( \begin{array}{cc} -1 & 0 \\ 0 & 1\end{array} \right).
\end{equation}
Let the mutually orthogonal qubit states be
\begin{eqnarray}
|\delta_{0}\rangle = X_{\delta}^{\dagger}|0\rangle = cos \delta |0\rangle + sin \delta |1\rangle \ , \nonumber \\
|\delta_{1}\rangle = X_{\delta}^{\dagger}|1\rangle = -sin \delta |0\rangle + cos \delta |1\rangle \ .
\end{eqnarray}
The operator $c_{1}\bar{R}_{uc,m} = I_{8N}-2|1,u_{c},m\rangle \langle 1,u_{c},m|$ is the negative of the reflection about $|1\rangle|u_{c}\rangle|m\rangle$ state. It is implemented by applying $\bar{R}_{uc,m}$ in the joint space iff the ancilla qubit is in $|1\rangle$ state. As $\bar{R}_{uc,m}$ can be implemented in one step, $c_{1}\bar{R}_{uc,m}$ also takes one step. Similarly, the controlled walk operator $c_{1}W$ performs a quantum walk $W$ in the joint space iff the ancilla qubit is in $|1\rangle$ state. 
Thus $U_{C}$ takes $2$ time steps for implementation, one for $c_{1}\bar{R}_{uc,m}$ and another for $c_{1}W$.

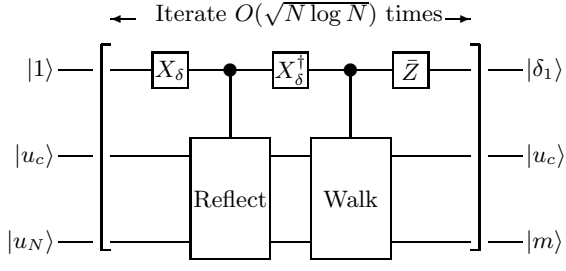
\begin{figure}
\setlength{\unitlength}{0.65pt}
\begin{picture}(160,160)(80,0)
\put(23,25){\makebox(0,0)[r]{$\vert{u_{N}}\rangle$}}

\put(297,25){\makebox(0,0)[l]{$\vert{m}\rangle$}}

\put(165,150){\makebox(0,0)[b]{Iterate $O(\sqrt{N\log N})$ times}}
\put(70,155){\vector(-1,0){15}}
\put(250,155){\vector(1,0){15}}

\put(25,25){\line(1,0){20}}
\put(25,75){\line(1,0){20}}
\put(25,125){\line(1,0){20}}

\put(275,125){\line(1,0){20}}

\put(275,25){\line(1,0){20}}
\put(275,75){\line(1,0){20}}
\put(240,125){\line(1,0){25}}

\put(80,115){\framebox(20,20){$X_{\delta}$}}

\put(150,115){\framebox(20,20){$X_{\delta}^{\dagger}$}}
\put(220,115){\framebox(20,20){$\bar{Z}$}}

\put(218,25){\line(1,0){47}}
\put(218,75){\line(1,0){47}}
\put(102,15){\framebox(46,70){\small{Reflect}}}

\put(148,25){\line(1,0){24}}

\put(148,75){\line(1,0){24}}

\put(195,125){\circle*{8}}
\put(195,121){\line(0,-1){36}}

\put(172,15){\framebox(46,70){Walk}}

\put(23,125){\makebox(0,0)[r]{$\vert{1}\rangle$}}
\put(23,75){\makebox(0,0)[r]{$\vert{u_{c}}\rangle$}}

\put(297,125){\makebox(0,0)[l]{$\vert{\delta_{1}}\rangle$}}
\put(297,75){\makebox(0,0)[l]{$\vert{u_{c}}\rangle$}}

\put(125,125){\circle*{8}}

\put(170,125){\line(1,0){50}}
\put(100,125){\line(1,0){50}}
\put(55,125){\line(1,0){25}}

\put(125,121){\line(0,-1){36}}

\put(55,25){\line(1,0){47}}

\put(55,75){\line(1,0){47}}
\put(270,20){\line(0,1){120}}
\put(50,20){\line(0,1){120}}
\put(265,20){\line(1,0){5}}
\put(265,140){\line(1,0){5}}
\put(50,20){\line(1,0){5}}
\put(50,140){\line(1,0){5}}

\end{picture}

\caption{Circuit diagram for the controlled quantum walk search algorithm. The Reflect and Walk boxes denote the reflection operator $\bar{R}_{uc,m}$ and the walk operator $W$ as defined in the text.} 
\end{figure}

For $\delta = 0$, the ancilla qubit is redundant and the new algorithm reduces to AKR's algorithm. The optimal algorithm is obtained if we choose $\delta$ such that $cos\delta = \Theta(\sqrt{1/\log N})$. Then measurement of the lattice state, after $O(\sqrt{N\log N})$ iterations of the operator $U_{C}$, gives the desired state $|m\rangle$ with constant probability. The total complexity of algorithm is therefore $O(\sqrt{N\log N})$.

To analyse the algorithm, we first show that the controlled quantum walk algorithm is an instance of the abstract search algorithm. We have $(X_{\delta}^{\dagger})_{b}\cdot c_{1}\bar{R}_{uc,m} \cdot (X_{\delta})_{b}  = c_{\delta 1}\bar{R}_{uc,m}$ where $c_{\delta 1}\bar{R}_{uc,m}$ applies $\bar{R}_{uc,m}$ in the joint space iff the ancilla qubit is in $|\delta_{1}\rangle$ state. Eq. (\ref{definitionUC}) then implies that 
\begin{equation}
U_{C} = C \cdot c_{\delta 1}\bar{R}_{uc,m}\ ,\ C = (\hat{Z})_{b}\cdot c_{1}W.
\end{equation}
Since $c_{\delta 1}\bar{R}_{uc,m} = I_{8N} - 2|\delta_{1},u_{c},m\rangle\langle \delta_{1},u_{c},m|$ is equivalent to $R_{\delta 1,uc,m} = 2|\delta_{1},u_{c},m\rangle\langle \delta_{1},u_{c},m|-I_{8N}$ upto a sign, the effective target state of the algorithm is 
\begin{equation}
|t_{\delta}\rangle = |\delta_{1}\rangle|u_{c}\rangle|m\rangle.
\end{equation}
We need to find the eigenspectrum of $C$ and the expansion coefficients of $|t_{\delta}\rangle$ in its eigenbasis. If $|\Phi\rangle_{J}$ is an eigenvector of the walk operator $W$ with eigenvalue $e^{i\theta}$ then it is easy to check that $|1\rangle_{b}|\Phi\rangle_{J}$ is an eigenstate of the operator $C$ with the same eigenvalue $e^{i\theta}$. Explicitly, 
\begin{equation}
|1\rangle_{b}|\Phi\rangle_{J}\stackrel{c_{1}W}{\rightarrow}e^{i\theta}|1\rangle_{b}|\Phi\rangle_{J}\stackrel{\bar{Z}}{\rightarrow}e^{i\theta}|1\rangle_{b}|\Phi\rangle_{J}. 
\end{equation}
Similarly, $|0\rangle_{b}|\Phi\rangle_{J}$ is an eigenvector of $C$ with the eigenvalue $-1$ due to the $\bar{Z}$ operator. Hence the subspace spanned by the states $|0\rangle_{b}|\psi\rangle_{J}$ is the $-1$-eigenspace of $C$. 
Having determined the eigenspectrum of $C$ in terms of that of $W$, we can easily infer that $C$ satisfies the required conditions for the abstract search algorithm within the subspace $\mathcal{H}_{b}\otimes \mathcal{H}_{0}$. Moreover, this subspace is preserved by the operator $U_{C}$.

To quantify the dynamics of the algorithm, we now calculate the quantities given by Eqns. (\ref{abstractexpansion}-\ref{abstracttargetoverlap}). Let $a_{0}(\delta)$, $a_{pq}(\delta)$ and $a_{k}(\delta)$ denote the expansion coefficients of the target state $|t_{\delta}\rangle$ in the eigenbasis of $C$. We have 
\begin{equation}
a_{pq}(\delta) = \langle \delta_{1},u_{c},m|1,\Phi_{pq}\rangle = a_{pq}cos\delta.
\end{equation}
where $a_{pq}$ are the expansion coefficients of $|u_{c},m\rangle$ in the eigenbasis of $W$, discussed in the previous section. Similarly, we get $a_{0}(\delta) = a_{0}cos\delta$.
Apart from these, the projection $A_{k}$ of $|t_{\delta}\rangle$ on the $-1$-eigenspace of $C$ is non-zero. It corresponds to the ancilla qubit being in $|0\rangle$ state, so $A_{k}(\delta)= |\langle \delta_{1}|0\rangle| = |sin\delta|$. This projection was not significant in AKR's algorithm, but it is crucial for the new algorithm. 

Using these values, and Eqns. (\ref{AKRsum1}--\ref{AKRsum3}) for the sums occuring in Eqns. (\ref{abstractexpansion}--\ref{abstracttargetoverlap}), we find that the two relevant eigenvectors of $U_{C}$ are $|\pm \alpha_{\delta}\rangle$ with the eigenvalues $e^{\pm i\alpha_{\delta}}$, with
\begin{equation}
\alpha_{\delta} = \Theta\left(\frac{1}{\sqrt{N\left(\log N + \frac{tan^{2}\delta}{4}\right)}}\right). \label{CQWeigenvalues}
\end{equation}
The overlap of the initial state $|\Phi_{0}\rangle_{cqw}$ with $|\alpha^{-}_{\delta}\rangle$ is
\begin{equation}
|\langle \alpha^{-}_{\delta}|\Phi_{0}\rangle_{cqw}| \geq 1-\Theta \left(\frac{1}{\log^{2}N}\right)-\Theta(N\alpha_{\delta}^{4}tan^{2}\delta). \label{CQWinitial}
\end{equation}
After $T = \lceil \frac{\pi}{4\alpha_{\delta}} \rceil$ iterations of $U_{C}$, we get the state $|\alpha^{+}_{\delta}\rangle$. Its overlap with the $|\delta_{1},u_{c},m\rangle$ state is
\begin{equation}
|\langle \delta_{1},u_{c},m|\alpha^{+}_{\delta}\rangle|= min\left(\Theta\left(\frac{1}{cos\delta\sqrt{\log N}}\right),1\right). \label{CQWtarget}
\end{equation} 

We consider the special case when $cos\delta = \Theta(\sqrt{1/\log N})$. In this case, the $|\alpha^{+}_{\delta}\rangle$ state has a constant overlap with the desired $|m\rangle$ state, and hence measuring the state will give $|m\rangle$ with constant probability. Using $tan^{2}\delta = \Theta(\log N)$ in Eq. (\ref{CQWeigenvalues}), we find that $\alpha_{\delta}=\Theta(1/\sqrt{N\log N})$. Putting it in Eq. (\ref{CQWinitial}), we get $|\langle \alpha^{-}_{\delta}|\Phi_{0}\rangle_{cqw}| = 1-\Theta(1/\log^{2}N)$ so the initial state is very close to $|\alpha_{\delta}^{-}\rangle$. The required number of iterations to get the state $|\alpha^{+}_{\delta}\rangle$ is $O(1/\alpha_{\delta}) = O(\sqrt{N\log N})$. Thus the time complexity of the algorithm is $O(\sqrt{N\log N})$.

If we choose $cos\delta \ll \sqrt{1/\log N}$, then using Eq. (\ref{CQWtarget}), we find that the $|\alpha^{+}_{\delta}\rangle$ state has still a constant overlap with the desired $|m\rangle$ state, but $tan^2\delta \gg \log N$ and the number of iterations required to get the $|\alpha^{+}_{\delta}\rangle$ state is much higher than $O(\sqrt{N\log N})$. If we choose $cos\delta \gg \sqrt{1/\log N}$, then the number of iterations required to get $|\alpha^{+}_{\delta}\rangle$ state remains $O(\sqrt{N\log N})$, but this state is no longer close to the desired state $|m\rangle$ and quantum amplitude amplification is needed to get to the desired state. The balance is achieved when $cos\delta = \Theta(\sqrt{1/\log N})$. 


\section{Discussion}
We have presented a modification of the discrete time quantum walk search algorithm by Ambainis, Kempe and Rivosh for the problem of two-dimensional spatial search. The new algorithm solves the problem in $O(\sqrt{N\log N})$ time steps and improves on AKR's algorithm by a factor of $O(\sqrt{\log N})$. It can be easily generalized to the continuous time quantum walk algorithm by Childs and Goldstone~\cite{CG}. In the continuous walk algorithm, the system is evolved under a time-independent Hamiltonian and the restriction on the Hamiltonian is that it should couple only neighboring sites. To apply the new algorithm, we just attach an ancilla qubit to the Hilbert space and then evolve the whole system under a suitably controlled Hamiltonian.

It is an open question that whether the performance of algorithm can be further improved. As the problem has a lower bound of $\Omega(\sqrt{N})$ time steps~\cite{optimal}, it will be interesting to get an algorithm which can solve the problem in $O(\sqrt{N})$ time steps or to show that no further improvement over $O(\sqrt{N\log N})$ complexity is possible. Within the framework considered here, probably $O(\sqrt{N\log N})$ complexity is the best that can be achieved. The minimum eigenvalue gap of the walk operator or the Hamiltonian is $O(1/\sqrt{N\log N})$, so the adiabaticity condition demands a minimum evolution time $O(\sqrt{N\log N})$. Even the algorithms of AKR and CG evolve the system for $O(\sqrt{N\log N})$ time, but their final statea are not close to the desired state. In the new algorithm, we have introduced \emph{extra} eigenstates of the walk operator by attaching an ancilla qubit. These extra eigenstates allow interference in such a way that the final state gets close to the desired state.   

The algorithm presented in this paper assumes a unique marked item, but it can be easily generalised to the case of multiple marked items with $O(\log N)$ overhead in computational complexity~\cite{aaronson}, making the total complexity of the algorithm $O(\sqrt{N}\log^{3/2}N)$. In their paper, AKR have extended their algorithm to the case of two marked items (see section 6.5 of ~\cite{AKR}), and they have shown that the algorithm succeeds in only $O(\sqrt{N}\log N)$ time steps for this case. The same extension applies to the new algorithm which solves the same problem in $O(\sqrt{N\log N})$ time steps. Similarly, the extension of AKR's algorithm to the case of two-dimensional coin-space (see Theorem 3 of~\cite{AKR}) also applies to the new algorithm.

Finally, we point out that the new algorithm can be applied to any instance of the abstract search algorithm, but the improvement factor may not be significant. In the case of higher-than-two dimensional spatial search, AKR's algorithm solves the problem in $c\sqrt{N}$ time steps where $c$ is a constant (see Theorem 4 of ~\cite{AKR}). By using the new algorithm, we can improve the performance only by a constant factor. It can be shown that if $c\gg 1$, then the performance can be improved by a factor of $\sqrt{c}$, making the total complexity $\sqrt{cN}$ (see section III.B of ~\cite{tulsi}). For $c=O(1)$, there is not much improvement, obviously because $\Omega(\sqrt{N})$ is the lower bound on any quantum search algorithm. 

{\it Note added:} After this work was completed, Prof. Apoorva Patel
pointed out to me that similar improvement in algorithm complexity can be
obtained using the Dirac equation with a mass term \cite{patel}.
A non-zero value for the mass eliminates the infrared divergence, and
provides the best performance when scaled appropriately with the lattice size.

\vspace{0.3cm}

\textbf{Acknowledgments}: I thank Prof. Apoorva Patel for going through the manuscript and for helpful comments and discussions. 

\vspace{0.4cm}

\section*{APPENDIX: ABSTRACT SEARCH ALGORITHM}
Here, we present the analysis of the abstract search algorithm, which iterates the operator $U_{A}=U\cdot R_{t}$ on the state $|\Phi_{0}\rangle$ that is a unique eigenstate of $U$ with eigenvalue $1$. Here $R_{t}$ is the reflection operator about the target state $|t\rangle$ and $U$ is required to be a real operator. The analysis closely follows that of AKR (see section $7$ of ~\cite{AKR}) with the difference that we have considered the possibility that $U$ may have an eigenspace with eigenvalue $-1$, referred to as the $-1$-eigenspace here. We will find the relevant features of the eigenspectrum of $U_{A}$, which are completely determined by the eigenspectrum of $U$ and the expansion coefficients of $|t\rangle$ in the eigenbasis of $U$. As discussed in section II, the target state $|t\rangle$ can be expanded in the eigenbasis of $U$ as
\begin{displaymath}
|t\rangle = a_{0}|\Phi_{0}\rangle +\sum_{j}a_{j}(|\Phi_{j}^{+}\rangle +|\Phi_{j}^{-}\rangle) +\sum_{k}a_{k}|\Phi_{k}\rangle.
\end{displaymath}
For convenience, we use the notations $a_{l}$, $|\Phi_{l}\rangle$ and $\theta_{l}$ ($l\in \{0,j,k\}$), for denoting the expansion coefficients $a_{0},a_{j},a_{k}$, the eigenvectors $|\Phi_{0}\rangle, |\Phi_{j}\rangle, |\Phi_{k}\rangle$, and the eigenangles $\theta_{0}=0,\theta_{j}\not\in\{0,\pi\},\theta_{k}=\pi$ respectively. 

We define for real $\lambda$, the unnormalised vector $|w_{\lambda}\rangle$, whose expansion coefficients in the eigenbasis of $U$ are given by $\langle\Phi_{l}|w_{\lambda}\rangle =a_{l}F_{\lambda}(\theta_{l})$, $F_{\lambda}(\theta_{l})=cot(\frac{\lambda-\theta_{l}}{2})$. We state some relations satisfied by function $F_{\lambda}$, which we will use later. These relations can be derived easily as is done in ~\cite{AKR}. 
\begin{eqnarray}
&e^{i\theta}(-1+iF_{\lambda}(\theta)) = e^{i\lambda}(1+iF_{\lambda}(\theta)),& \\
&F_{\lambda}(\theta)+F_{\lambda}(-\theta) = \frac{2sin \lambda}{cos\theta-cos \lambda},& \\
&F_{\lambda}(0)=cot\frac{\lambda}{2}\ ;\ F_{\lambda}(\pi) = -tan\frac{\lambda}{2}.& 
\end{eqnarray}

As shown by AKR, if $|w_{\lambda}\rangle$ is orthogonal to $|t\rangle$ then the unnormalised vector $|\lambda\rangle =|t\rangle + i|w_{\lambda}\rangle$ is an eigenvector of the operator $U_{A}=U\cdot R_{t}$ with the eigenvalue $e^{i\lambda}$. It is because of the special properties of the function $F_{\lambda}(\theta_{l})$. To see this, we note that the expansion coefficients of $|\lambda\rangle$ in the eigenbasis of $U$ are
\begin{equation} 
\langle \Phi_{l}|\lambda\rangle =\langle \Phi_{l}|t\rangle + i\langle \Phi_{l}|w_{\lambda}\rangle = a_{l}(1+iF_{\lambda}(\theta_{l})).
\end{equation}
 We have $R_{t}|\lambda\rangle = -|t\rangle + i|w_{\lambda}\rangle$ as $R_{t}$ does not alter $|w_{\lambda}\rangle$, orthogonal to $|t\rangle$ by assumption. Hence we have 
\begin{equation}
\langle \Phi_{l}|R_{t}|\lambda\rangle =-\langle \Phi_{l}|t\rangle + i\langle \Phi_{l}|w_{\lambda}\rangle = a_{l}(-1+iF_{\lambda}(\theta_{l})).
\end{equation} 
Since $|\Phi_{l}\rangle$ are eigenvectors of $U$, we have $\langle \Phi_{l}|U\cdot R_{t}|\lambda\rangle = e^{i\theta_{l}}\langle \Phi_{l}|R_{t}|\lambda\rangle = a_{l}e^{i\theta_{l}}(-1+iF_{\lambda}(\theta_{l}))$. Using Eqns. (24,27), we find it to be equal to $\langle \Phi_{l}|U_{A}|\lambda\rangle =e^{i\lambda}a_{l}(1+iF_{\lambda}(\theta_{l})) = e^{i\lambda}\langle \Phi_{l}|\lambda\rangle$. As this holds for all the basis vectors $|\Phi_{l}\rangle$, we find that $|\lambda\rangle$ is an eigenvector of $U_{A}=U\cdot R_{t}$ with eigenvalue $e^{i\lambda}$, iff $|w_{\lambda}\rangle$ is orthogonal to $|t\rangle$. This condition is equivalent to $\sum_{l}a_{l}^{2}F_{\lambda}(\theta_{l})=0$. Expanding this sum for $l=0,j$ and $k$, and using Eq. (25) for the term $F_{\lambda}(\theta_{j})+F_{\lambda}(-\theta_{j})$ occuring in the sum, we find the condition to be  
\begin{equation}
a_{0}^{2}\frac{cot\frac{\lambda}{2}}{sin \lambda}=\sum_{j}\frac{2a_{j}^{2}}{cos \lambda-cos\theta_{j}}+A_{k}^{2}\frac{tan\frac{\lambda}{2}}{sin \lambda},
\end{equation}
where $A_{k}=\sqrt{\sum_{k}a_{k}^{2}}$ is the projection of $|t\rangle$ state on $-1$-eigenspace of $U$.
It is easy to check that if above equation is satisfied for $\lambda$ then it is also satisfied for $-\lambda$ and vice versa.

Let $\theta_{min}$ be the smallest of $\theta_{j}$. Then as shown by AKR, the above equation has exactly two solutions, $\lambda = \alpha$ and $\lambda = -\alpha$, such that $|\alpha| < \theta_{min}/2$. Moreover, the eigenvectors corresponding to these eigenvalues are relevant as $|\Phi_{0}\rangle$ is almost completely spanned by them, and hence iteration of $U_{A}$ on $|\Phi_{0}\rangle$ can be analysed by considering only these eigenvectors. Typically $\theta_{min}$ is very small (in the case of two-dimensional spatial search, it is $O(1/\sqrt{N})$, and therefore $\alpha$ is very small. Writing above equation upto first order in $\alpha$, we get 
\begin{equation}
\frac{a_{0}^{2}}{\alpha^{2}}=\sum_{j}\frac{a_{j}^{2}}{cos\alpha-cos\theta_{j}}+\frac{A_{k}^{2}}{4}.
\end{equation} 
As shown by AKR, the first term on R.H.S. is $\Theta(\sum_{j}\frac{a_{j}^{2}}{1-cos\theta_{j}})$, which leads to
\begin{equation}
\alpha= \Theta \left(\frac{a_{0}}{\sqrt{\sum_{j}\frac{a_{j}^{2}}{1-cos\theta_{j}}+\frac{A_{k}^{2}}{4}}}\right).
\end{equation}

Let $|\pm\alpha\rangle = |t\rangle + i|w_{\pm \alpha}\rangle$ be the unnormalised eigenvectors of $U_{A}$ corresponding to the eigenvalues $e^{\pm i\alpha}$. Let $|\alpha^{-}_{u}\rangle = |\alpha\rangle - |-\alpha\rangle = i(|w_{\alpha}\rangle - |w_{-\alpha}\rangle)$ be an unnormalised state and let $|\alpha^{-}\rangle = |\alpha^{-}_{u}\rangle/\|\alpha^{-}_{u}\|$ be the corresponding normalized state. To show that the initial state $|\Phi_{0}\rangle$ is spanned by the eigenvectors $|\pm \alpha\rangle$, we find the overlap of $|\Phi_{0}\rangle$ with the vector $|\alpha^{-}\rangle$. The expansion coefficients of the vector $|\alpha^{-}_{u}\rangle$ in the eigenbasis of $U$ are given by 
\begin{equation}
|\langle\Phi_{l}|\alpha^{-}_{u}\rangle| = \langle\Phi_{l}|w_{\alpha}\rangle - \langle\Phi_{l}|w_{-\alpha}\rangle = a_{l}(F_{\alpha}(\theta_{l})-F_{-\alpha}(\theta_{l})).
\end{equation}
We have $|\langle\Phi_{0}|\alpha^{-}\rangle| = \frac{|\langle\Phi_{0}|\alpha^{-}_{u}\rangle|}{\|\alpha^{-}_{u}\|}$. Putting $l=0$ in above equation, we find $|\langle\Phi_{0}|\alpha^{-}_{u}\rangle|=a_{0}(F_{\alpha}(0)-F_{-\alpha}(0))=2a_{0}cot\frac{\alpha}{2}$, and hence we need to bound $\|\alpha^{-}_{u}\|$ to bound $|\langle\Phi_{0}|\alpha^{-}\rangle|$. Now
\begin{equation}
\|\alpha^{-}_{u}\| = \sqrt{\sum_{l}|a_{l}(F_{\alpha}(\theta_{l})-F_{-\alpha}(\theta_{l}))|^{2}}.
\end{equation}
In the summation over $l$, the term $T_{0}$ corresponding to $l=0$ is equal to $T_{0}= |\langle\Phi_{0}|\alpha^{-}_{u}\rangle|^{2} = 4a_{0}^{2}cot^{2}\frac{\alpha}{2} = \Theta(a_{0}^{2}/\alpha^{2})$. Similarly the term $T_{k}$ corresponding to $l \in k$ is equal to $4A_{k}^{2}tan^{2}\frac{\alpha}{2} = \Theta(A_{k}^{2}\alpha^{2})$. The term $T_{j}$ corresponding to $l \in j$ was calculated by AKR and found to be $T_{j}=\Theta(\alpha^{2}\sum_{j}a_{j}^{2}/(1-cos\theta_{j})^{2})$. Moreover, in the case of spatial search, they have shown that $T_{j}$ and $T_{k}$ are small compared to $T_{0}$ for large $N$. Hence, using $\|\alpha^{-}_{u}\| = \sqrt{T_{0}+T_{j}+T_{k}}$, we get $|\langle\Phi_{0}|\alpha^{-}\rangle| = \sqrt{T_{0}}/\|\alpha^{-}_{u}\| = 1-\frac{T_{j}+T_{k}}{2T_{0}}$. More explicitly
\begin{equation}
 |\langle \Phi_{0}|\alpha^{-}\rangle| \geq 1-\Theta \left(\alpha^{4}\sum_{j}\frac{a_{j}^{2}}{a_{0}^{2}}\frac{1}{(1-cos\theta_{j})^{2}}\right) - \Theta\left(\frac{A_{k}^{2}\alpha^{4}}{a_{0}^{2}}\right).
\end{equation}

Thus the state $|\Phi_{0}\rangle$ is very close to $|\alpha^{-}\rangle = c(|\alpha\rangle - |-\alpha\rangle)$, where $c$ is the normalization factor. As $|\pm \alpha\rangle$ are the eigenvectors of $U_{A}$ with eigenvalues $e^{\pm i\alpha}$, we have $(U_{A})^{q}|\alpha^{-}\rangle = c(e^{iq\alpha}|\alpha\rangle - e^{-iq\alpha}|-\alpha\rangle)$. After $T=\lceil \pi/2\alpha \rceil$ iterations of $U_{A}$, we come very close to the state $|\alpha^{+}\rangle = c(|\alpha\rangle + |-\alpha\rangle)$. 

The last part of analysis is to calculate the overlap between $|t\rangle$ and $|\alpha^{+}\rangle$ states. Let $|\alpha^{+}_{u}\rangle = |\alpha\rangle +|-\alpha\rangle$ be an unnormalized state. We have $|\alpha^{+}\rangle = |\alpha^{+}_{u}\rangle/\|\alpha^{+}_{u}\|$ and hence $|\langle t|\alpha^{+}\rangle| = \frac{|\langle t|\alpha^{+}_{u}\rangle|}{\|\alpha^{+}_{u}\|}$. As $|\alpha^{+}_{u}\rangle = 2|t\rangle + i(|w_{\alpha}\rangle +|w_{-\alpha}\rangle)$ and $|w_{\pm \alpha}\rangle$ are orthogonal to $|t\rangle$, we find $|\langle t|\alpha^{+}_{u}\rangle|$ to be equal to $2$. Similarly,
\begin{equation}
\|\alpha^{+}_{u}\|^{2} = \|2|t\rangle + i(|w_{\alpha}\rangle +|w_{-\alpha}\rangle)\|^{2} = 4+\|w_{\alpha}+w_{-\alpha}\|^{2}.
\end{equation}
The expansion coefficients of the vector $|w_{\alpha}+w_{-\alpha}\rangle$ in the eigenbasis of $U$ are given by $\langle\Phi_{l}|w_{\alpha} + w_{-\alpha}\rangle = a_{l}(F_{\alpha}(\theta_{l})+F_{-\alpha}(\theta_{l}))$, and hence
\begin{equation}
\|w_{\alpha} +w_{-\alpha}\|^{2} = \sum_{l}|a_{l}(F_{\alpha}(\theta_{l})+F_{-\alpha}(\theta_{l}))|^{2}.
\end{equation}
For $l \in \{0,k\}$, the term $F_{\alpha}(\theta_{l})+F_{-\alpha}(\theta_{l})$ vanishes as $\theta_{l}$ is either $0$ or $\pi$ for such $l$'s and $F_{\alpha}(\theta_{l}) = -F_{-\alpha}(\theta_{l})$ for $\theta_{l} \in \{0,\pi\}$. So, all the non-vanishing terms in above sum correspond to $l\in j$. This sum has been computed by AKR and shown to be $\Theta(\sum_{j}a_{j}^{2}cot^{2}\frac{\theta_{j}}{4})$. Putting it in Eq. (35), we get
\begin{equation}
|\langle t|\alpha^{+}\rangle| = \left(1+\Theta \left(\sum_{j}a_{j}^{2}cot^{2}\frac{\theta_{j}}{4}\right)\right)^{-1/2}.
\end{equation}
When the sum is large compared to $1$ (as in the case of AKR's spatial search algorithm), we get
\begin{equation}
|\langle t|\alpha^{+}\rangle|= \Theta \left(min\left(\frac{1}{\sqrt{\sum_{j}a_{j}^{2}cot^{2}\frac{\theta_{j}}{4}}},1\right)\right).
\end{equation}
This completes the analysis of the abstract search algorithm.

\end{document}